\documentclass[a4paper,prd,twocolumn,nofootinbib,preprintnumbers]{revtex4}

\usepackage{graphicx}

\begin{document}

\title{Searching for New Physics through LFV Processes
\footnote{Talk given at the 2nd International Conference on Flavor Physics,
KIAS, Seoul, Korea, October 6-11, 2003.}}

\author{Yasuhiro Okada}
\email{yasuhiro.okada@kek.jp}
\affiliation{Theory Group, KEK, Tsukuba, Ibaraki 305-0801, Japan}
\affiliation{Department of Particle and Nuclear Physics,
             The Graduate University of Advanced Studies,
             Tsukuba, Ibaraki 305-0801, Japan}

\begin{abstract}
Muon lepton flavor processes are reviewed in connection with search for
physics beyond the standard model. Several methods to distinguish 
different theoretical models are discussed for $\mu \to e \gamma$, $\mu  \to 3 e$, 
and $\mu - e$ conversion processes. New calculation of
the $\mu - e$ conversion rate is presented including a Higgs boson 
mediated effect in the supersymmetric seesaw model.
\end{abstract}


\preprint{KEK-TH-942}

\maketitle
\section{Introduction}
Search for lepton flavor violation (LFV) in charged lepton processes is one of 
promising ways to look for physics beyond the standard model. 
In the standard model without neutrino masses, the lepton number is
conserved separately for each generation, so that the process like
$\mu \to e \gamma$ is strictly forbidden. LFV search has been carried out 
since the early days of muon experiments, and experimental upper bounds 
have been improved continuously by about two orders of magnitude per a decade. 
In fact, the absence of the $\mu \to e \gamma$ process was a motivation to 
introduce the second neutrino, and therefore the generation structure in the 
particle physics.

Recent discovery in neutrino oscillations suggests that the lepton flavor
is not strictly conserved in Nature. However, expected branching ratios
for charged lepton LFV processes are much suppressed in the simplest extension
of the standard model which accommodates the neutrino oscillation, namely
the seesaw model of neutrino mass or the Dirac neutrino model. 
The discovery of LFV in the charged lepton sector therefore implies 
existence of new particles and new interactions beyond the simple
seesaw model, most likely in the energy scale close to the electroweak
scale. 

Supersymmetry(SUSY) is a very important example of new physics which can be explored 
by LFV processes. SUSY models introduce SUSY partners of leptons,
namely sleptons, and the mass matrixes of the slepton can be sources of LFV.
In fact, since the low-energy SUSY model was considered in early 80's, it has 
been noted that LFV processes put severe constraints on the flavor off-diagonal
terms of the slepton mass matrixes. In recent years, searches for LFV processes 
has attracted more attention. This is because it was pointed out that 
the predicted branching 
ratios of LFV processes can be close to the present experimental upper bounds for
some of well-motivated SUSY models such as SUSY GUT and the SUSY seesaw model
\cite{Kuno:1999jp}.
Unlike the non-SUSY version of the seesaw neutrino model, it is possible that the 
neutrino Yukawa coupling constant can be sources of both neutrino oscillations and 
sizable LFV in the charged lepton sector. 

In this talk, muon LFV processes are reviewed in the first part. In particular, 
discussions are given on how we can distinguish different theoretical models. 
In the second part,  a
new calculation on the $\mu - e$ conversion rate is presented including a 
Higgs boson mediated effect in the SUSY seesaw model. This is an example illustrating usefulness of observables discussed in the first part. 
 
\section{Muon LFV processes}
There are three important LFV processes related to muon, namely, $\mu \to e \gamma$, 
$\mu  \to 3 e$, and $\mu - e$ conversion in a muonic atom. Current experimental bounds 
on these processes as well as tau LFV processes are listed in Table 1.
\begin{table}[htbp]
 \caption{Experimental bounds of LFV processes}
 \begin{center}
  \begin{tabular}{|c|c|}
    \hline
    Processes  & Current bound \\
    \hline
    $\mu^+ \to e^+ \gamma $ & $1.2 \times 10^{-11} $ \\
    \hline
    $\mu^+ \to e^+e^+e^- $  & $1.0 \times 10^{-12}$ \\
    \hline
    $\mu^- Ti \to e^- Ti $ & $6.1 \times 10^{-13}$ \\
    \hline
    $\tau \to \mu \gamma$  & $3.1 \times 10^{-7}$  \\
    \hline
    $ \tau \to \mu \eta $  &  $3.4 \times 10^{-7}$ \\
    \hline
    $ \tau \to lll$  &  $1.4 - 3.1 \times 10^{-7}$  \\
    \hline
  \end{tabular}
 \end{center}
\end{table}
The upper bounds on tau LFV processes in this table
are taken from new results of the Belle experiment\cite{belle}. 
In $ \tau \to lll$, $l$ represents electron or muon, so that the listed values 
corresponds to the range of upper bounds for various combinations
of electrons and muons. As we can see in this table, the upper bounds
are stronger for muon processes than tau processes. Relationship between
tau and muon LFV processes depends on a theoretical model under consideration.
For examples, muon processes severely constrain the tau
LFV processes in the SUSY seesaw model, although there is some parameter
space where the branching ratios of tau LFV processes are closer to the 
experimental bounds.    

Among three muon processes, searches of $\mu\to e \gamma $ and  $\mu \to 3e $
processes are carried out by positive muon decays, whereas negative muons 
are used in the $\mu-e$ conversion experiment. In the former case, positive muons
are stopped in a target, and experimental signature is simultaneous emissions of 
$e^+$ and $\gamma$, or  $e^+$, $e^+$, and $e^-$ with appropriate kinematics. 
On the other hand, a negative muon behaves like a heavy electron in matter, 
so that it falls into the 1s atomic orbit soon after it is stopped in a target.
The negative muon eventually either is captured in a nucleus by weak interaction
emitting a neutrino, or decays in an orbit. If LFV interaction exists, muon can be
converted to electron without a neutrino emission. In the $\mu-e$ conversion 
experiment, this electron emission is searched for. There are two possibilities
about final states of the nucleus. If the nucleus state remains to be the same 
grand state as the initial state, the energy of the electron is monochromatic. 
If the nuclear state is excited, the electron can have a broad spectrum. 
In general, the grand state to
grand state transition is dominant, because this is a coherent process, and 
the transition probability has, roughly speaking, an enhancement factor of the atomic 
number. The bound listed in Table 1 corresponds to the coherent transition.

There are future plans for $\mu\to e \gamma $ and $\mu-e$ conversion experiments.
In the MEG experiment which is under constriction at PSI, 
the $\mu\to e \gamma $ will be 
searched to the level of $10^{-14}$\cite{PSI}. For $\mu-e$ conversion, 
the MECO experiment at BNL
is planed to cover $10^{-16}$ and below for aluminum target \cite{MECO}. In future, 
further improvement of $\mu-e$ conversion search by two orders of magnitudes 
is discussed as a part of 
a future muon facility at J-PARC (PRIME experiment) \cite{PRISM}. 

If LFV is discovered in these experiments, the next step is to determine
the nature of new interaction so that we can distinguish various theoretical models.
I will discuss several ways to obtain insights on new interactions.    

\subsection{Comparison of three muon processes}
The effective Lagrangian describing the $\mu \to e\gamma$ transition is given by
\begin{eqnarray}
{\cal L}_{\rm photon}&=& 
- \frac{4 G_{\rm F}} {2^{1/2}} \left(
m_{\mu}A_R \overline{\mu_R}\sigma ^{\mu \nu} e_L F_{\mu \nu} \right.
\nonumber\\
&&
\left. +m_{\mu}A_L \overline{\mu_L}\sigma ^{\mu \nu} e_R F_{\mu \nu}+h.c. \right).
\label{photon}
\end{eqnarray}  
$A_R$ and $A_L$ are coupling constants, which are functions of parameters of models
including LFV interactions. The above operators also make contributions to $\mu \to 3e$ 
and $\mu -e$ conversion processes. On the other hand, the latter two processes can 
depend on other type of operators, namely, $\bar{\mu}e \bar{e} e$ and 
$\bar{e}\mu\bar{q}q$ type four-fermion interactions, respectively. 
In many interesting cases including some of SUSY models, however, 
the photon penguin operators give dominant contributions in other two processes.
In such cases we can derive relations among the branching ratios of three processes.
\begin{eqnarray}
B(\mu^+ \to e^+e^+e-) &\sim& 6\times10^{-3}B(\mu\to e\gamma),
\label{3e-rel}\\
\frac{\sigma(\mu^- Ti \to e^- Ti)}{\sigma(\mu^- Ti \to capture)}&\sim& 4 \times 10^{-3}
B(\mu\to e\gamma).
\label{mue-rel}
\end{eqnarray}
These relations are important to classify types of new physics
models with LFV interactions.
\subsection{Muon polarization in $\mu \to e\gamma $ and $\mu \to 3e$ decays}
If muons are polarized, angular distributions with respect to the initial
muon polarization provide important information on nature of LFV interactions.

For $\mu \to e\gamma $, the angular distribution is given as follows:
\begin{eqnarray}
   \frac{dB(\mu^+\to e^+\gamma)}{d\cos{\theta}}&\propto &
   1+\frac{|A_L|^2-|A_R|^2}{|A_L|^2+|A_R|^2}P_{\mu}\cos{\theta},
\end{eqnarray}
where $P_{\mu}$ is the muon polarization and the $\theta$ is the angle between 
the muon polarization direction and the positron momentum. The parity asymmetry
( $A_{\mu \to e\gamma}=(|A_L|^2-|A_R|^2)/(|A_L|^2+|A_R|^2)$) is particularly 
important in SUSY models, because this depends on whether the flavor mixing 
exists in the left-handed slepton sector, the right-handed slepton sector, or
both. For example, $A_{\mu \to e\gamma}=-1$ for the SUSY seesaw model without GUT
because only left-handed slepton sector has the flavor mixing. On the other hand
GUT models in general have non-zero values for both $A_L$ and $A_R$.    

For $\mu \to 3e$, we can define two P-odd asymmetries and one T-odd asymmetry
\cite{Okada:1997fz}. The T-odd asymmetry is defined as the up-down asymmetry of 
the initial muon polarization direction with respect to the final decay plane. 
In the SU(5) SUSY GUT, the T-odd asymmetry can be as large as 15\% if we include the
CP phase in the left-right mixing term in the slepton mass matrix. On the other
hand, in the SO(10) SUSY GUT, the T-odd asymmetry is small. This is because the 
photon penguin operators in Eq.(\ref{photon}) give dominant contributions to
the $\mu \to 3e$ amplitude, whereas interference between the photon penguin
and four-fermion operators are necessary to generate the T-odd quantities.
In this case, however, two P-odd asymmetries of $\mu \to 3e$ have 
definite relations with the $\mu \to e \gamma$ asymmetry $A_{\mu \to e\gamma}$.
Together with the branching ration relation in Eg.(\ref{3e-rel}), these 
relations provide evidence that the LFV interaction is dominated by the 
photon-penguin operators.  
   
\subsection{Atomic number dependence of $\mu-e$ conversion branching ratio}
In $\mu -e$ conversion experiments, choice of appropriate target nucleus is 
an important issue. Although actual planning of experiments involves many technical
aspects, estimation of backgrounds, etc, basic physical input is the atomic number ($Z$)
dependence of the $\mu-e$ conversion branching ratios for a given Lagrangian at the 
quark and lepton level. We therefore calculated the coherent $\mu -e$ conversion 
rates in various nuclei for general LFV interactions \cite{Kitano:2002mt}. As shown 
below, the $Z$-dependence provides another way to distinguish various theoretical models.

The quark level Lagrangian relevant for coherent $\mu-e$ conversion processes are given
as follows:
\begin{eqnarray}
    {\cal L}_{\rm int} &=&
    - \frac{4 G_{\rm F}}{2^{1/2}}
    \left(
    m_\mu A_R \bar{\mu} \sigma^{\mu \nu} P_L e F_{\mu \nu}
    \right.
   \nonumber \\
   &&
    \left.
    + m_\mu A_L \bar{\mu} \sigma^{\mu \nu} P_R e F_{\mu \nu}
    + {\rm h.c.}
    \right) \nonumber \\
    &&
    - \frac{G_{\rm F}}{2^{1/2}} \sum_{q = u,d,s}
    \left[ \right.
    \nonumber \\
    &&
    \left(
    g_{LS(q)} \bar{e} P_R \mu + g_{RS(q)} \bar{e} P_L \mu
    \right) \bar{q} q  \nonumber \\
    && 
    +
    \left(
    g_{LV(q)} \bar{e} \gamma^{\mu} P_L \mu
    + g_{RV(q)} \bar{e} \gamma^{\mu} P_R \mu
    \right) \bar{q} \gamma_{\mu} q \nonumber \\
    &&
	+ {\rm h.c.}
    \left. \right].
    \label{1}
\end{eqnarray}
There are three contributions in the above Lagrangian. The first one is
"dipole" operator, which is the same one responsible to the $\mu \to e \gamma$
decay. The other two are "scalar" and "vector" operators, namely the quark 
currents are scalar and vector types, respectively. Each of three contributions
have two operators according to the structure of the lepton current.

In the calculation of the $\mu-e$ conversion branching ratio, we used a fully 
relativistic formalism with the most updated nuclear data. The relativistic effect
turned out to be very important in heavy nuclei. Although the charged
density in various nuclei is very precisely determined, the neutron density 
is not very well-known and becomes a source of ambiguity. In \cite{Kitano:2002mt},
detail discussion on the theoretical ambiguity is given.

\begin{figure}[t]
    \begin{center}
	\includegraphics[width=8cm]{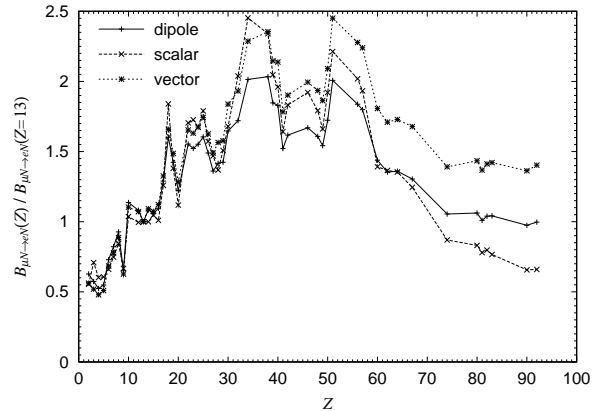} 
    \end{center}
    \caption{
The $\mu$-$e$ conversion ratios for the typical theoretical models are
plotted as functions of the atomic number $Z$.  The solid, the long
dashed, and the dashed lines represent the cases that the photonic
dipole, scalar, and vector operator dominates, respectively. In the calculation, 
an approximation is used where nomailzed proton and neutron distributions 
are taken to be the same. The conversion ratios are normalized by the 
conversion ratio in aluminum nuclei ($Z =13$).
} \label{okada-f1}
\end{figure}

In Fig.\ref{okada-f1}, the $Z$-dependence of the $\mu-e$ conversion rate is
shown for three types operators. The conversion branching ratio, namely
the transition probability divided by the capture probability, is normalized by
its value for $Al$, and the resulting factor is shown in this figure.
We can see that the branching ratio is largest for atomic numbers of
30-60 for any of three types of operators. Since the $\mu -e$ conversion
process is a coherent process and the capture process is not, we expect 
that the branching ratio is larger for large nuclei. This is true up to
a certain nucleus, but not for heavy nuclei. The $\mu -e$ conversion amplitude
is essentially given by the overlapping integral among three quantities, 
the initial muon 1s wave function, the nucleon density, and the final electron 
wave function. For heavy nuclei, the finite size effect becomes important
in this overlapping integral and acts as a suppression factor, so that the
conversion branching ratio is maximum in the intermediate nuclei.
We also note that there is little difference in Z-dependence in lighter
nuclei for different operators, but sizable difference appears for heavy
nuclei. This is due to a relativistic effect of the muon wave function. In fact 
the conversion formula reduces to the same without a relativistic effect.

We can therefore use this dependence of the $\mu-e$ conversion rate as a mean
to discriminate different models. For examples, 
$B(\mu Pb \to e Pb)/B(\mu Al \to e Al)$= 1.0, 0.77, 1.4 for dipole, scalar and vector
operators.

\section{Higgs-mediated $\mu-e$ conversion in the SUSY seesaw model}
One of reasons why LFV search has attracted much attention in recent years 
is that a sizable branching ratio is predicted in well-motivated cases
of SUSY models. In particular, the SUSY seesaw model
has a new source of flavor mixing in the neutrino Yukawa coupling constant.
Through the renormalization due to the Yukawa coupling constant, the flavor 
off-diagonal terms are induced in the slepton mass matrixes. The resulting 
branching ratios of  LFV processes are large if the Yukawa coupling constant
is sizable. Although there is not exact one-to-one correspondence between
the neutrino mixing and the LFV processes, the branching ratios of muon LFV processes
can be close to the experimental bounds, if the right-handed neutrino mass scale is
$10^{13} -10^{14}$ GeV. 

In this section, we discuss a new contribution to the $\mu-e$ conversion process
in the SUSY model due to exchange of the Higggs boson. Although we take a SUSY seesaw 
model as an explicit example, this effect can be important for other cases, 
as long as we consider a large value
of $\tan{\beta}$, which is a ratio of two vacuum expectation values.

The Higgs sector of the SUSY model consists of two Higgs doublets. 
This is required to write down all relevant Yukawa coupling constants in a SUSY
invariant way. Namely, one Higgs doublet $(H_1)$ couples to down-type quarks 
and charged leptons, and the other $(H_2)$ couples to up-type quarks. 
The Higgs sector is so called 
"Type II" two Higgs doublet model. This is, however, the structure of Yukawa coupling
at the tree level. If there are sizable corrections due to SUSY particle loop diagrams,
it is possible that the effective Yukawa coupling is not exactly in the form of
the Type II model, but of a general two Higgs doublet model. The new contribution is in particular important when we consider a particle spectrum where SUSY particles 
are much heavier than all Higgs bosons. In such a case, the effective theory below the
SUSY mass threshold is a general two Higgs doublet model. 
Many interesting new effects are pointed out in various flavor changing neutral 
current processes as well as LFV processes\cite{FCNC}. For LFV processes, 
new Higgs-boson exchange contributions are considered for $\tau \to 3 \mu$ 
\cite{Babu:2002et,Dedes:2002rh}
and $\tau \to \mu \eta$ \cite{Sher:2002ew} processes.

We calculated the $\mu-e$ conversion rate in the SUSY seesaw model, taking 
account of the Higgs-mediated contribution \cite{Kitano:2003wn}.
Due to the SUSY loop correction to the Yukawa coupling constant, new couplings 
between $H_2$ and charged leptons are induced, which include
LFV interactions. Then, the heavy neutral Higgs boson exchange diagram generates the
LFV interaction of the $\bar{e_L}\mu_R \bar{s}s$ form. After taking the matrix elements
between nucleons, this operator can give dominant contributions to the $\mu-e$ 
conversion rare, especially for a large $\tan{\beta}$ region, because this contribution
has a $(\tan{\beta})^6$ dependence. Roughly speaking the Higgs-exchange contribution
is given by
\begin{eqnarray}
 B(\mu {\rm Al} \to e {\rm Al})_{H^0} \sim O(10^{-13}) \cdot 
\left( \frac{200 {\rm GeV}}{m_{H^0}} \right)^4 \cdot
\left( \frac{\tan \beta}{60} \right)^6 ,
\end{eqnarray}
whereas the ordinary photon exchange contribution is 
\begin{eqnarray}
B(\mu {\rm Al} \to e {\rm Al})_\gamma \sim 
O(10^{-13}) \cdot \left( \frac{1000 {\rm GeV}}{M_S} \right)^4
\cdot 
\left(
\frac{\tan\beta}{60}
\right)^2 .
\end{eqnarray}
  
\begin{figure}[t]
\begin{center}
\includegraphics[width=8cm]{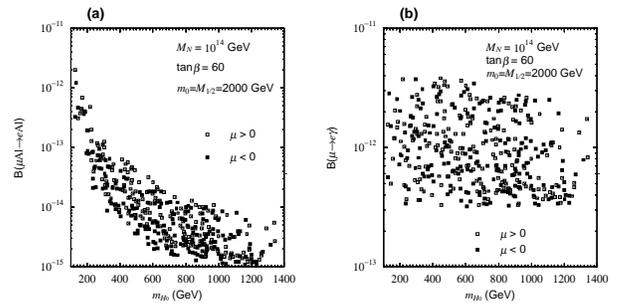} 
\end{center}
\caption{
The $m_{H^0}$ dependences of the branching ratios of the following
processes are shown: (a) $\mu-e$ conversion in aluminum nucleus and (b)
$\mu \to e \gamma$ decay.
We take the right-handed neutrino masses to be $10^{14}$ GeV, and $\tan
\beta = 60$.  The soft masses for the Higgs fields are treated as free
parameters.  } 
\label{okada-f2}
\end{figure}

\begin{figure}[t]
\begin{center}
\includegraphics[width=8cm]{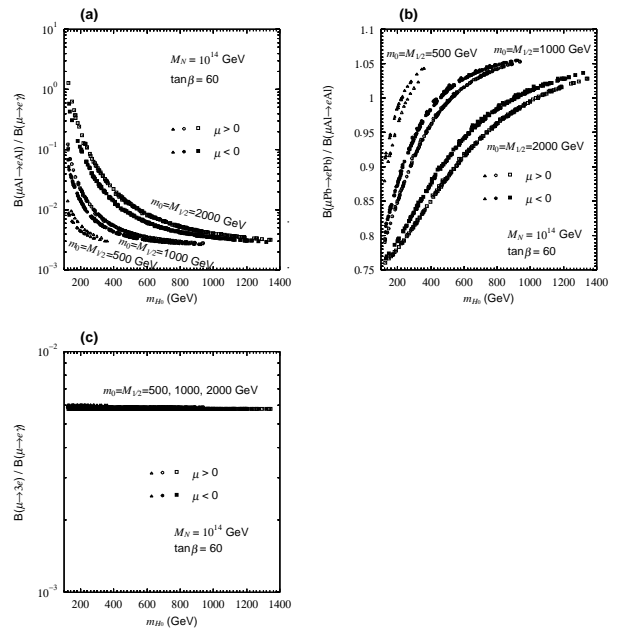} 
\end{center}
\caption{
The following ratios of the branching ratios are shown as functions of
$m_{H^0}$: (a) $B(\mu {\rm Al} \to e {\rm Al}) / B(\mu \to e \gamma)$,
(b) $B(\mu {\rm Pb} \to e {\rm Pb}) / B(\mu {\rm Al} \to e {\rm Al})$,
and (c) $B(\mu \to 3e) / B(\mu \to e \gamma)$.
We take the right-handed neutrino masses to be $10^{14}$ GeV, and $\tan
\beta = 60$.  The soft masses for the Higgs fields are treated as free
parameters.  } 
\label{okada-f3}
\end{figure}

In Figs. \ref{okada-f2} and \ref{okada-f3}, results of numerical calculations are shown.
Here, we take the right-handed neutrino mass of $10^{14}$ GeV and $\tan{\beta}=60$.
For more details on calculation, see Ref.\cite{Kitano:2003wn}.
Fig. \ref{okada-f2} shows the $\mu-e$ conversion rate is enhanced if the heavy Higgs boson
mass ($m_{H_0}$) becomes smaller, whereas the $\mu \to e\gamma$ process 
does not show any particular
dependence on $m_{H_0}$. For a smaller $m_{H_0}$, both processes can have 
branching ratios of $10^{-13} - 10^{-12}$. 
The enhancement for the $\mu-e$ conversion process is shown in Fig.\ref{okada-f3},
where the ratio of two branching ratios are plotted. For a large $m_{H_0}$ the ratio 
approaches  to the value of the photon-penguin dominant case, whereas the ratio
can be close to one for a small $m_{H_0}$. The dominance of the Higgs-mediated 
contribution can be seen by the atomic number dependence of the $\mu-e$ conversion
rate. In the region of the $\mu-e$ conversion enhancement,
$B(\mu Pb \to e Pb)/B(\mu Al \to e Al)$ becomes smaller, which is realized 
when the scalar coupling gives a dominant contribution.
We also show $B(\mu \to 3e)/B(\mu \to e \gamma)$ here. There is no sizable enhancement
for $\mu \to 3e$ process because the Higgs-exchange diagram involves a small electron
Yukawa coupling constant.

\section{Conclusions}
In this talk, I discussed searches for new physics effects in muon LFV processes,
namely, $\mu \to e \gamma$, $\mu \to 3e$ and $\mu-e$ conversion processes.
Several ways are considered to discriminate different theoretical models. 
Comparison of three branching ratios is one important way, but within a single
process, we can obtain information on the LFV interaction using muon polarization
for  $\mu \to e \gamma$ and $\mu \to 3e$ processes and Z-dependence of the conversion
rate for  the $\mu-e$ conversion case. As an illustration, we have calculated
the Higgs-mediated contribution to  $\mu-e$ conversion in the SUSY seesaw model, 
and shown how
the ratio of the branching ratio and Z-dependence  are useful to discriminate 
different models. Search for LFV processes is therefore a very powerful method to look
for new physics, and if LFV effects are discovered, we expect rich physics program
in all of these muon processes.  

This work was supported
in part by a Grant-in-Aid of the Ministry of Education, Culture, Sports,
Science, and Technology, Government of Japan (No.~13640309).

\end{document}